\documentclass[twocolumn,aps,prl,showpacs,preprintnumbers,footinbib,floatfix,superscriptaddress]{revtex4}
\usepackage[english]{babel}
\usepackage{pdfpages}
\usepackage{amsmath}
\usepackage{amssymb}
\usepackage{color}
\usepackage{graphicx}
\newcommand{\lf}{\left}
\newcommand{\rg}{\right}
\newcommand{\be}{\begin{equation}}
\newcommand{\ee}{\end{equation}}
\newcommand{\bea}{\begin{eqnarray}}
\newcommand{\eea}{\end{eqnarray}}
\newcommand{\nn}{\nonumber}
\newcommand{\ba}{\begin{array}}
\newcommand{\ea}{\end{array}}

\renewcommand{\a}{\alpha}
\renewcommand{\b}{\beta}

\renewcommand{\d}{\delta}
\newcommand{\ve}{\varepsilon}

\renewcommand{\o}{\omega}

\newcommand{\s}{\sigma}
 \newcommand{\la}{\langle}
 \newcommand{\ra}{\rangle}
 \newcommand{\br}{{\bf r}}
 \newcommand{\bk}{{\bf k}}
  \newcommand{\bp}{{\bf p}}
\newcommand{\bq}{{\bf q}}
\def\pd{\partial}    
\def\lb{\label}
\def\bE{{\bf E}}
\def\bv v{{\bf v}}

\def\bJ{{\bf J}}

\def\pref#1{(\ref{#1})}

\begin{document}
%

\title{Unconventional dc transport in Rashba electron gases}

\author{Valentina Brosco}
\affiliation{ISC-CNR and Department of Physics, Sapienza University of Rome, P.le A. Moro 2, 00185 Rome, Italy}
\author{Lara Benfatto}
\affiliation{ISC-CNR and Department of Physics, Sapienza University of Rome, P.le A. Moro 2, 00185 Rome, Italy}
\author{Emmanuele Cappelluti}
\affiliation{ISC-CNR and Department of Physics, Sapienza University of Rome, P.le A. Moro 2, 00185 Rome, Italy}
\author{Claudio Grimaldi}\affiliation{Laboratory of Physics of Complex Matter, Ecole Polytechnique F\'ed\'erale de
Lausanne, Station 3, CH-1015 Lausanne, Switzerland}

\begin{abstract}
We discuss the transport properties of a disordered two-dimensional
electron gas with  strong Rashba spin-orbit  coupling.
We show that in the high-density regime where
the Fermi energy overcomes the energy
associated with spin-orbit coupling, dc transport is accurately
described by a standard Drude's law,
due to a non-trivial compensation between the suppression of
back-scattering and the relativistic correction to the quasi-particle
velocity.
On the contrary, when the system enters the opposite
\emph{dominant} spin-orbit regime, Drude's paradigm breaks down
and the dc conductivity becomes strongly sensitive to the spin-orbit
coupling strength,  providing a suitable tool to test the
entanglement between spin and charge degrees of freedom in these
systems.\end{abstract}
\maketitle

Spin-orbit (SO) coupling is a fundamental ingredient in
spintronics\cite{awschalom2007} as it provides an advantageous locking between spin and electron orbital momentum.
Recently, intense research efforts \cite{manchon2014} have been devoted to  two-dimensional (2D) materials with broken
inversion symmetry, where the
SO strength, parametrized by a characteristic energy scale, $E_0$, can be tuned
by means of external conditions (electric fields, gating, doping, pressure, strain, \ldots).
In most of these systems (ex.: surface alloys
\cite{ast2007,ast2008,gierz2009,mirhosseini2010,yaji2010,sanchez2013,gruznev2014},
layered bismuth tellurohalides
\cite{eremeev2012,bahramy2012,sakano2013,xi2013,chen2013,ye2015,xiang2015},
HgTe quantum wells \cite{gui2004},
interfaces  between complex oxides
\cite{ohtomo2004,reyren2007,caviglia2008,bell2009,benshalom2010,biscaras2012,seri2012,joshua2013,liu2013,hurand2015,zhong2015,caviglia2010,zhong2013,biscaras2014,joshua2012})
the total charge carrier density $n$ can be tuned
down to very small concentrations, implying very small Fermi energies $E_{\rm F}$.
Although the high-density (HD) regime $E_{\rm F}\gtrsim E_0$
has been widely investigated\cite{khaetskii_prl06,raimondi2001,schwab2002,raimondi2005,argwal_prb11,bercioux2015,manchon2014}, relatively less attention has been paid to the
opposite regime of dominant SO (DSO), $E_0 \gtrsim E_{\rm F}$.

In this Letter we provide a detailed investigation of the dc
conductivity  of a 2D electron gas (2DEG) with Rashba\cite{bychkov1984} SO coupling
in the different density regimes. Using a Boltzmann approach and
a fully quantum analysis based on Kubo formula,
we show that in the high-density  regime $E_{\rm F}\gtrsim E_0$,
dc transport is {\em independent} of the SO strength, and the dc conductivity $\sigma_{dc}$
of electrons having effective mass $m$ and scattering time $\tau_0$ follows
the conventional Drude law for 2DEGs,
\bea
\sigma_{\rm Drude}
&=&
\frac{n e^2 \tau_0}{m},
\label{sdrude}
\eea
that results from a \emph{non-trivial} cancellation of the SO coupling effects on
the quasiparticle velocity and transport scattering time. Remarkably, as
soon as the system enters the DSO regime $E_0 \gtrsim E_{\rm F}$,
 Drude's paradigm \eqref{sdrude} breaks down and the dc
conductivity accurately follows the analytical formula:
\bea
\sigma_{\rm DSO }
&=&
\frac{e^2 \tau_0 n_0}{2\, m}\lf(\frac{n^4}{n_0^4}+\frac{n^2}{n_0^2}\rg)
\quad n\leq n_0,
\label{sigmasso}
\eea
where $n_0=2mE_0/(\pi\hbar^2)$ is the density at $E_F=E_0$.
In contrast to the linear dependence of $\sigma_{dc}$ on the charge density found
in the HD regime, $n\geq n_0$,
Eq. (\ref{sigmasso}) predicts an {\em unconventional} non-linear behavior
of $\sigma_{dc}$ with $n$, that is controlled by the SO interaction encoded in $n_0$.
The relevance of this result is twofold: demonstrating that dc transport is strongly sensitive to Rashba SO coupling,
not only it suggests that SO coupling could be measured in a transport experiment but also, what is more important
for applications, it points at the possibility of tuning the conductivity of a 2DEG by tuning the SO coupling
strength {\sl via} external gates.
%
%
%

A disordered Rashba 2DEG confined to the $(x,y)$-plane is described by the following Hamiltonian
\be
H=\!\! \int \!\!d{\bf r}\,\Psi^{\dag}({\bf r})\!\lf[\frac{ p^2}{2 m}+\a\, \hat z\, \cdot ({\bf p} \times \vec \sigma)+V_{\rm imp}({\bf r})\rg]\!\Psi({\bf r}) \label{model},\ee
where
$V_{\rm imp}({\bf r})$ is the disorder potential, $\alpha$ is
the SO coupling,
$\vec \sigma$ is the vector of Pauli matrices, $\Psi(\br)$ and $\Psi^\dag(\br)$
are spinor fields which respectively create  and destroy electrons at position $\br$ and $\hat a=\vec a/|\vec a|$.
Here we limit ourselves to the simplest case of  Gaussian random disorder with ``white noise'' correlations,
namely we set $\la V_{\rm imp}({\bf r})V_{\rm imp}({\bf r}')\ra_{\rm imp}=n_i v_{\rm imp}^2 \delta({\bf r}-{\bf r'})$
where $v_{\rm imp}$  and $n_i$ denote respectively  the scattering strength and the impurity density.
\begin{figure}[t]
\begin{center}
\includegraphics[width=8cm,clip=true]{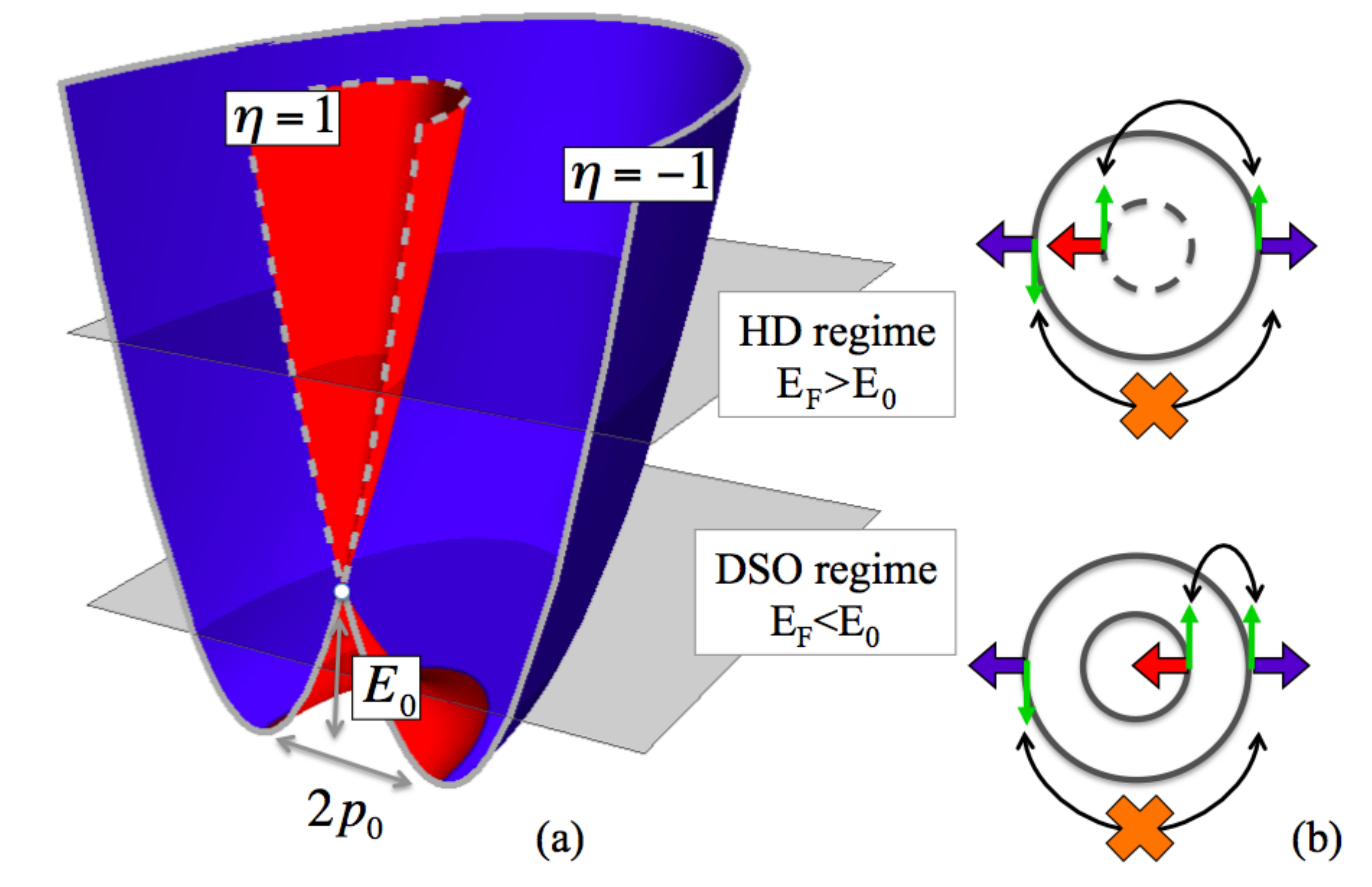}
  \caption{(Color online) (a):  dispersion of the Rashba model. The solid/dashed contours and the red/blue colours denote respectively $s$ and $\eta$. (b):  Examples of allowed and forbidden back-scattering processes. Thick (red and blue)  and thin green arrows denote  the quasi-particle velocity $\vec v_{{\bf p}s}$ and the spin $\la\vec\sigma\ra_{{\bf p}s} $ of each state, respectively.}\label{fig1}
  \end{center}
\end{figure}
In the absence of impurities, $H$ gives an electronic spectrum, depicted in Fig.\ref{fig1}, consisting
of two bands with dispersion  $E^s_p-E_0=(p+s p_0)^2/(2m)-E_0$, where $p_0=m\a$, $E_0=m\alpha^2/2$
and $s=\pm 1$ denotes the eigenvalue of the helicity operator, $S$, defined as usual as $S=\hat z\, \cdot (\hat p \times \vec \sigma)$.
In the following we measure the Fermi energy, $E_F$, from the lower band edge, so that  $E_F=E_0$ corresponds to the ``Dirac point'', $p=0$,
and the HD and DSO regimes are realised respectively for   $E_F>E_0$ and $E_F<E_0$.

As already noted in \cite{cappelluti2007}, the onset of the DSO regime leads to a rather sharp change in density of states (DOS) at the Fermi level. Indeed, while for $E_F>E_0$ the DOS  is constant, $N(E_F)=N_0=m/(\pi\hbar^2)$,  for $E_F<E_0$ it
displays a van Hove singularity, $N(E_F)=N_0\sqrt{E_0/E_F}$ leading to a modification of the dependence of the density on the Fermi energy \cite{cappelluti2007}:
\be
n\simeq \label{nwd}
\left\{
\begin{array}{ll}
 N_0(E_F+E_0) & E_F>E_0   \\[0.1cm]
2N_0 \sqrt{E_F E_0} & E_F<E_0.\ \\
\end{array}
\right.
\ee
%
In the presence of static diluted disorder, the singular behavior of the DOS
reflects directly on the quasiparticles lifetime that, using
Fermi golden rule (see {\sl e.g.} Ref.\cite{ziman}), can be cast as
$\tau(E_p^s)={\cal V}\,[\sum_{{\bf  p }'s'} Q^{{\bf p} s}_{{\bf p'} s'}]^{-1}$, where ${\cal V}$ is the 2D  volume of
the sample and
\be
\label{kernel}
Q^{{\bf p} s}_{{\bf p}' s'}=\pi n_i v_{imp}^2(1+s s' \hat p \cdot \hat p')\delta(E_p^s-E_{p'}^{s'}),
\ee
is the scattering kernel.
Using Eqs. (\ref{nwd}-\ref{kernel}) one can easily show that  the  quasi-particle scattering time  scales linearly  with the density  in the DSO regime, {\sl i.e.}
\be
\tau=\tau(E_F)= \label{tauqp}
\left\{
\begin{array}{ll}
\tau_0 & E_F>E_0   \\[0.1cm]
\tau_0 \sqrt{E_F/E_0}=\tau_0 (n/n_0) & E_F<E_0,\ \\
\end{array}
\right.
\ee
where $\tau_0=\hbar^2/(mn_i v_{imp}^2)$ denotes the quasi-particle scattering time in the absence of SO.
%

To explain the behavior of the conductivity across the different
regimes, we start by recalling the definition of the velocity
operator in the helicity basis,
\be\label{jlongitudinal-transverse} \lf[\vec {\rm v}\rg]_{ss'}=\vec v_{{\bf p}s}\delta_{ss'}-i\alpha s(1-\delta_{ss'})\hat t_p ,
\ee%
 where $\vec v_{{\bf p}s}=\nabla_{\bf p} E^s_p=\hat p(p/m+s\alpha)$
 denotes the quasi-particle velocity and $\hat t_p$ is defined as
 $\hat t_p=\{ p_y/p, - p_x/p\}$.
An important thing to underline here, general for any  chiral system,
is that in the presence of SO coupling also the velocity acquires a spin structure,
which has a deep impact on the transport properties.
%
%
%
%
As we discuss in more details below, to a first approximation the conductivity can be described within a standard semiclassical Boltzmann approach that only keeps  the quasiparticle current, arising from the diagonal components of the velocity operator \eqref{jlongitudinal-transverse}.  In the relaxation time approximation  at $T=0 $,  $\sigma_{dc}$  can be then estimated as %
\be
\sigma_{dc} \simeq\sigma^{\rm B}_{dc}=\frac{e^2}{2\cal V}\sum\nolimits_{{\bf p}s}\delta(E_F-E_{p}^s)|\vec v_{{\bf p}s}|^2\tau_{ps}^{\rm tr}\label{conboltz}
\ee
where the transport scattering times $\tau_{ps}^{\rm tr}$ satisfy
the following  equations
 \cite{Supp}  %
 \be
\frac{\tau_{ps}^{\rm tr}}{\tau(E^s_p)}=1+\frac{1}{\cal V}\sum_{{\bf p'}s'} Q^{{\bf p'}s'}_{{\bf p }s}\frac{ \vec v_{{\bf p}'s'}\cdot \vec v_{{\bf p}s} }{|\vec v_{{\bf p}s}|^2}
\tau_{p's'}^{\rm tr}.\label{tau}\ee
Using explicitly the definition of $Q^{{\bf p'}s'}_{{\bf p }s}$ \eqref{kernel}, we introduce the transport helicity index $\eta=s(\hat v_{{\bf p}s}\cdot \hat p)\equiv\hat z\cdot(\vec v_{{\bf p}s}\times \la\vec\sigma\ra_{{\bf p}s})=\pm 1$, that accounts for the reciprocal orientation of spin and velocity, and we recast Eq.\eqref{tau} as follows:
 \be\label{taueta}
\frac{\tau_{\eta}^{\rm tr}}{\tau}=1+\frac{n_i v_{\rm imp}^2}{4 v_F\hbar^2}\sum_{\eta'} \eta\eta' p_{\eta'} \tau^{\rm tr}_{\eta'},
\ee
%
%
where $v_F=\sqrt{2mE_F}$ and $p_\eta=|m v_F-\eta\, p_0|$ are the Fermi momenta on the inner and outer Fermi surfaces.
The above equation suggests that the index $\eta$   can be used to efficiently classify the states at the Fermi level across the different regimes. In particular, as illustrated in Fig.\ref{fig1}a, where  the value of $\eta$ is indicated by the red/blue colors of the surface, at $E_F>E_0$ $\eta$ simply coincides with $s$, on the contrary, at $E_F<E_0$  $\eta$ allows to distinguish between the two Fermi circles  that have the same value of $s$ but antiparallel quasi-particle velocities.
Using this classification $\sigma^{\rm B}_{dc}$ can be cast as
\be
\sigma^{\rm B}_{dc}=\frac{e^2 v_F}{4\pi}\sum\nolimits_{\eta}\tau^{\rm tr}_\eta p_\eta=\sum\nolimits_{\eta}\sigma_\eta, \label{conboltz2}
\ee
where the transport scattering times $\tau^{\rm tr}_\eta$ are given by the solution of Eqs. \eqref{taueta}
\be\label{tautr}
\tau^{\rm tr}_{\eta}=\tau\, p_\eta/\bar p_F,
\ee
%
%
%
with $\bar p_F=1/2\sum_\eta p_\eta$, {\sl i.e.} $\bar p_F=mv_F$  for $E_F>E_0$ and $\bar p_F=p_0$  for $E_F<E_0$.  
As one can easily check, Eqs.\eqref{conboltz2} and  \eqref{tautr} yield  Drude result \eqref{sdrude} at $E_F>E_0$, and Eq.(\ref{sigmasso}) in the DSO regime.

  The physical relevance of $\eta$, as  compared to the standard helicity, $s$, is evident in Fig.\ref{fig1}b, where we show that spin conservation forbids back-scattering between states having the same value of $\eta$. What is more important, in the DSO regime the only allowed back-scattering processes reverse the sign of the quasi-particle velocity without changing the direction of momentum.
These effects determine the  density dependence of the scattering times (a) and of the conductivities (b) of the the majority  ($\eta=-1$)  and minority  ($\eta=+1$)  carriers shown  in Fig.\ref{figboltz}. As one can see in Fig.\ref{figboltz}b, transport is in general dominated by the majority carriers that, due to the suppression of backscattering  have also the larger transport scattering time, $\tau_-^{\rm tr}>\tau>\tau_+^{\rm tr}$.
Let us focus on the transport properties of majority and minority carriers  across the different regimes.
%
In the HD regime $\tau$ is a constant and $\tau_-^{\rm tr}/\tau$ increases as the density decreases due to the shrinking of the inner Fermi circle. At $n=n_0$, where only states with $\eta=-1$ are present, back-scattering is completely suppressed and one recovers $\tau_-^{\rm tr}=2\tau$, like e.g. in graphene \cite{graphene_review}. However, differently from graphene, as long as $n>n_0$ a compensation between SO effects on the velocity and on the transport scattering times  of the two types of carriers restores the usual Drude conductivity, even for $n$ very close to $n_0$.
This result is non trivial: indeed,  setting  naively $\tau^{\rm tr}_{\eta}=\tau$  in Eq.\eqref{conboltz}, would lead to \cite{Supp} 
$\sigma\simeq (n-n_0/2)\tau_0/m$ {\sl i.e.} $\sigma<\sigma_{\rm Drude}$ even at  $n>n_0$ \cite{marsigliosr13}. 
On the other hand, as the system enters the DSO regime $\tau$ starts to decrease linearly as predicted by Eq.\ \eqref{tauqp} and backscattering processes for the majority carriers are progressively restored.
Both these effects quench $\tau_-^{\rm tr}$ as $n<n_0$, see Fig.\ \ref{figboltz}a, leading to an overall sublinear behavior of the conductivity, see Fig.\ \ref{figboltz}b.

%

\begin{figure}[t]
\centering
\includegraphics[trim=0cm 0cm 0cm 0cm, clip=true, width=8cm]{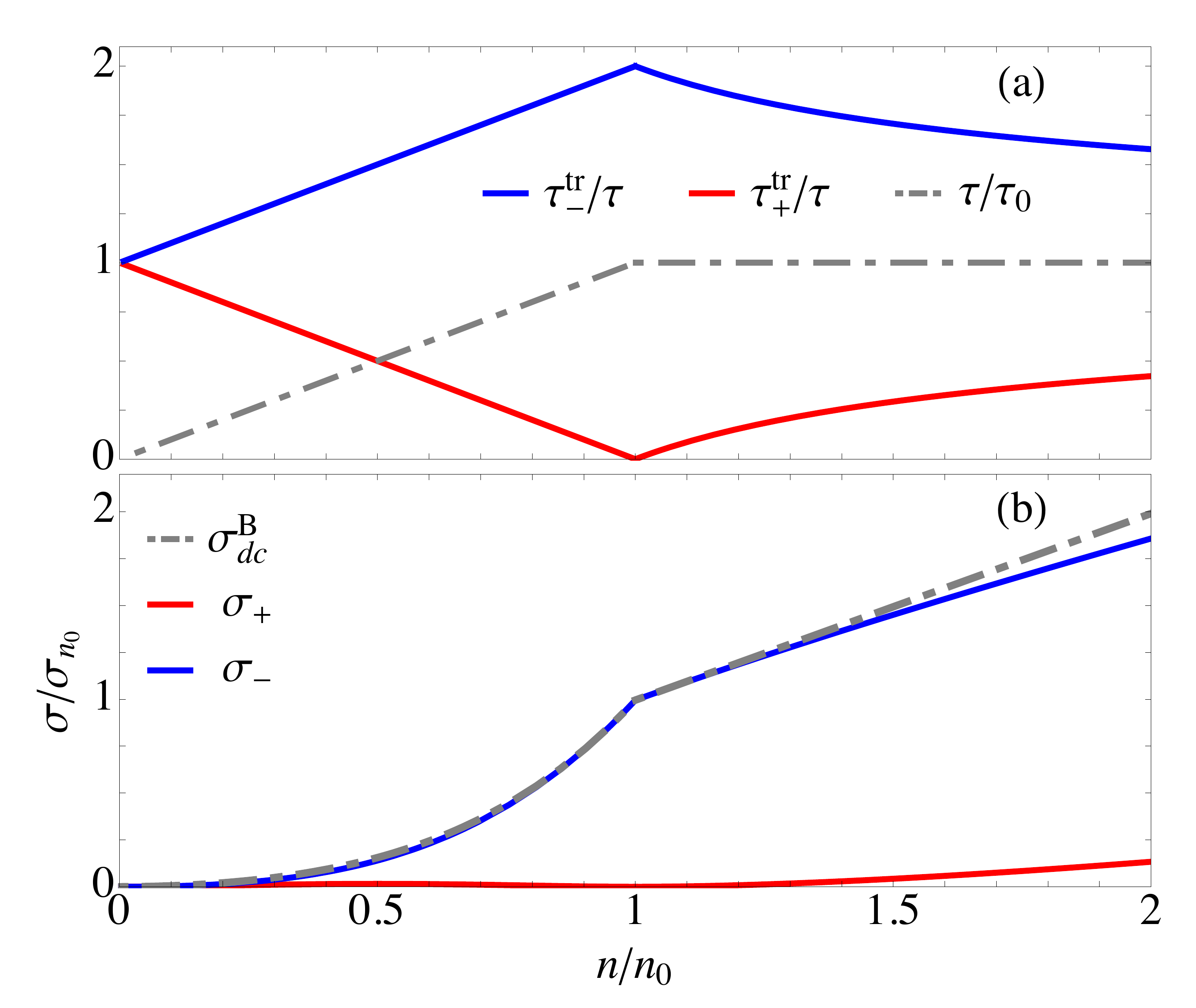}
  \caption{(Color online)  (a) Density dependence of the ratio $\tau_{\mp}/\tau$  for the majority ($\tau_-^{\rm tr}$, solid blue line) and minority ($\tau_+^{\rm tr}$, solid red line) carriers, and of the ratio $\tau/\tau_0$ (dashed line).
 (b) Density dependence of Boltzmann dc conductivity $\sigma^B_{dc}$, and contribution $\sigma_\mp$ of the two types of carriers, in units of $\sigma_{n_0}=n_0e^2\tau_0/m$. }\label{figboltz}
\end{figure}

A deeper insight on  dc transport comes from the generalisation of  Boltzmann approach
to fully include quantum effects. To this end, we use Kubo linear
response theory \cite{MahanManyParticlePhysics}. We start by noticing that, within the self-consistent
Born approximation (SCBA), the retarded Green's function is diagonal in the helicity basis and it is given by the following matrix \cite{MahanManyParticlePhysics,Grimaldi2006},
\begin{equation}
\label{green1}
\lf[G^R(\mathbf{p},\omega)\rg]_{ss'}
=g^R_s(p,\omega)\delta_{ss'},
\end{equation}
where  $g_s(p,\omega)=\left[\omega-E_p^s+E_F-\Sigma^R(\omega)\right]^{-1}$ denotes the
Green function of electrons with helicity $s$. The  self-energy $\Sigma^R(\o)=n_i v_{\rm imp}^2/(2{\cal V})
\sum\nolimits_{{\bf p},s}g_s^R(p,\o)$ is spin and momentum independent \cite{MahanManyParticlePhysics}.
At zero frequency its imaginary part defines the elastic scattering rate of quasiparticles,
$\Gamma=-{\rm Im}[\Sigma^R(0)]={n_i v_{\rm imp}^2 \pi}/(2{\cal V})
\sum\nolimits_{{\bf p},s} {\cal A}_{s}(p)$,
where ${\cal A}_s(p)=-(1/\pi){\rm Im}g_s^R(p,\o=0) $ is the spectral function of each helicity band. In Fig.\ref{fig-sigmamu}a we plot the numerical self-consistent $\Gamma$  as a function of the ratio $n/n_0$  for different values of the SO coupling and we compare it with Boltzmann result $\Gamma_B=1/(2\tau)$.
As expected Boltzmann result is accurate for large  $n/n_0$, where  $\Gamma\ll
E_F$ holds and one can approximate the spectral function
as ${\cal A}_{s}(p)\simeq \delta(E_p^s-E_F)$,
and its accuracy increases with increasing $E_0/\Gamma_0$. 
On the contrary, as $E_F$ approaches the band edge, the DOS singularity is smeared by disorder and finite-band effects cut-off the divergence of  Boltzmann result $\Gamma_B=\Gamma_0\sqrt{E_0/E_F}$ following from \eqref{tauqp}. In the extreme diluted limit, (shaded region in Fig.\ref{fig-sigmamu}a) $\Gamma$ as given by the SCBA vanishes  and the diffusive approximation breaks down \cite{Supp}.

Within  linear response theory the conductivity is given by Kubo formula and, at $T=0$, it is related to the on-shell current-current response function as follows \cite{MahanManyParticlePhysics}
\be\label{sigmaxx}
\sigma_{dc}=\frac{\hbar}{2\pi} \lf(P_{xx}^{AR} -{\rm Re}[P_{xx}^{RR}]\rg),
\ee
where $P_{xx}^{LM}$ is given by
\be P_{xx}^{LM}=\frac{1}{\cal V}\sum\nolimits_{\bf p} {\rm Tr}\left[j_x(\mathbf{p})G^L(\mathbf{p},0) J^{LM}_x(\mathbf{p})G^M(\mathbf{p},0)\right],\label{plm}\ee
and the superscripts $L,M=A,R$ indicate  advanced/retarded
quantities.  In the above equation  $j_x(\mathbf{p})$ and $J^{LM}_x(\mathbf{p})$ denote the bare and dressed currents and they are in general represented by $2\times2$ matrices in the helicity space. In particular, $ j_x(\mathbf{p})=e{\rm v}_x$ is proportional to the
bare velocity \eqref{jlongitudinal-transverse}, while $ J_x$
has to be determined self-consistently \cite{Supp} and
 it can be written as
%
\be\lf[ \vec J^{LM}(\mathbf{p})\rg]_{ss'}= e \lf[\vec V^{LM}_{{\bf p}s}\delta_{ss'}-i\tilde\alpha^{LM}s(1-\delta_{ss'})\hat t_p\rg]\label{JLM},
\ee
where $\vec V^{LM}_{{\bf p}s}={\bf p}/m+s\tilde \alpha^{LM} \hat p$
denotes the dressed quasi-particle velocity. By comparing Eqs.\ \eqref{jlongitudinal-transverse} and \eqref{JLM} one sees that, as usual \cite{MahanManyParticlePhysics} in Kubo formalism the effects of scattering by impurities are encoded, {\sl via} the vertex function $\tilde\alpha^{LM}$, in the renormalization of the velocity. As we show below, under appropriate conditions, these effects are equivalently accounted for in Boltzmann language by the transport scattering times.
%
%

The anomalous velocity $\tilde\alpha^{LM}$ also plays an important role in the spin-Hall effect \cite{dyakonov1971}.
In this context it was shown that, although $\tilde\alpha^{RA}=0$ in the HD regime \cite{raimondi2005,raimondi2011,vignale2009} and $\tilde \alpha^{RA}\neq 0$ in the DSO regime  \cite{Grimaldi2006}, in both regimes the spin-Hall conductivity vanishes. This  result follows straightforwardly from the vanishing of $\tilde\alpha^{RA}$ in the HD regime while it can be proven by an explicit calculation in the DSO regime \cite{Grimaldi2006}.

 Using Eqs.\eqref{jlongitudinal-transverse} and \eqref{JLM}, the current response function \eqref{plm} can be cast as the sum
of inter- and intra-band terms: $P_{xx}^{LM}= P_{\rm intra}^{LM}+P_{\rm inter}^{LM}$, where
\bea
P_{\rm intra}^{LM}&=&\frac{e^2}{2{\cal V}} \sum\nolimits_{\bf p s} \vec v_{{\bf p}s}
\cdot \vec {\rm V}^{LM}_{{\bf p}s} g^L_s(p,0)g^M_s(p,0)\label{ALMV} \\
 P_{\rm inter}^{LM}&=&\frac{e^2}{2{\cal V}} \alpha\, \tilde \alpha^{LM}
 \sum\nolimits_{{\bf p}\, s\neq s'}g^L_s(p,0)g^M_{s'}(p,0).\label{BLMV}
\eea
From a numerical self-consistent solution of the self-energy and vertex equations, we calculate the fully quantum dc conductivity Eq.\eqref{sigmaxx}.
The results are shown in Fig. \ref{fig-sigmamu}(b), where we plot the conductivity  as a function of the electronic density  and of the SO coupling. Here we also plot Boltzmnann conductivity (blue lines) for two values of $E_0$, showing that  Kubo results follow quite closely  Boltzmann prediction.


The equivalence between the two approaches can be proven in the the limit of vanishingly small broadening of the spectral functions, $\Gamma\ll E_F$, where we can discard\cite{MahanManyParticlePhysics} the $RR$ term in Eq. \eqref{sigmaxx}. Indeed by neglecting also the inter-band contribution to $P^{RA}$,  relevant only at $n\simeq n_0$ where the spectral functions of the two chiral bands overlap, we can recast the conductivity as
\be
\label{app}
\sigma_{dc}\simeq\frac{e^2}{4{\cal V}\Gamma}  \sum\nolimits_{\bf p s} \vec v_{{\bf p}s}\cdot  \vec V^{RA}_{{\bf p}s} \delta(E_F-E_p^s),
\ee
%
%
that, by direct comparison with \eqref{conboltz},  yields $\sigma_{dc}\simeq\sigma^B_{dc}$ provided that $\vec V^{RA}_{{\bf p}s}\simeq\tau^{tr}_{{\bf p}s}\vec v_{{\bf p}s}/\tau$.
In the limit $\Gamma\ll E_F$ this relation is a straightforward consequence of the vanishing of $\tilde\alpha^{RA}$ for $E_F>E_0$ and it can be easily proved  for $E_F<E_0$ using $\tilde \alpha^{RA}\simeq\alpha(1-E_F/E_0)$ \cite{Supp}. This shows in particular that on the Fermi circles $V_{{\bf p}s}/v_F=\tau^{\rm tr}_\eta/\tau$.
%

The deviations between Boltzmann and Kubo results are better seen in  Fig. \ref{fig-sigmamu}c,
where we compare the corresponding mobilities, respectively defined as   $\mu_B=\sigma^{\rm B}_{dc}/(en)$ and $\mu_t=\sigma_{dc}/(en)$, and they can be ultimately ascribed to two factors.
First,  finite-band effects, that are mostly relevant for $n\lesssim 0.3n_0$ and  
are responsible for the deviations of $\Gamma$ from $\Gamma_B$, shown in Fig.\ \ref{fig-sigmamu}a, and for the relevance of $RR$ terms \cite{Supp} that  in turn imply that  for small densities  $\mu_t$ tends to saturate, in contrast to $\mu_B$.
Second,  inter-band terms that are mostly relevant at $n\simeq n_0$ and give a smoothening of dependence of $\mu_t$ on $n$.
This effect could be also captured by replacing the semiclassical Boltzmann equation with a fully quantum kinetic equation that includes also the off-diagonal  components of the velocity operator \eqref{jlongitudinal-transverse}  and of the non-equilibrium density-matrix in the helicity space. This allows one to account, in the presence of external fields, for the coherent superpositions of states with different helicities, as explained {\sl e.g.} in Refs. \cite{dyakonov84,khaetskii_prl06,shytov2006,raimondi2006}.

\begin{figure}[t]
\begin{center}
\includegraphics[width=8cm]{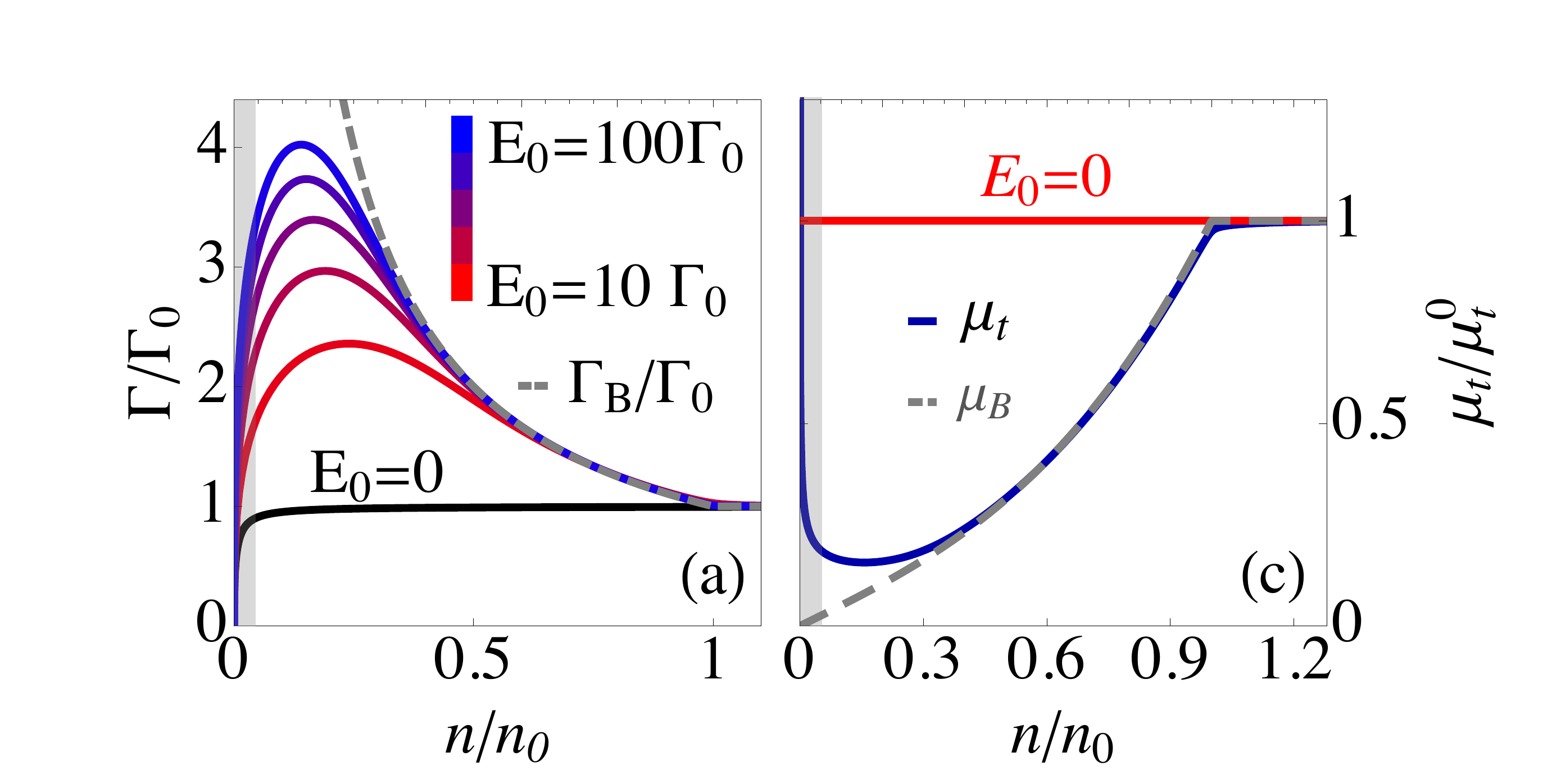}
\includegraphics[width=0.45\textwidth]{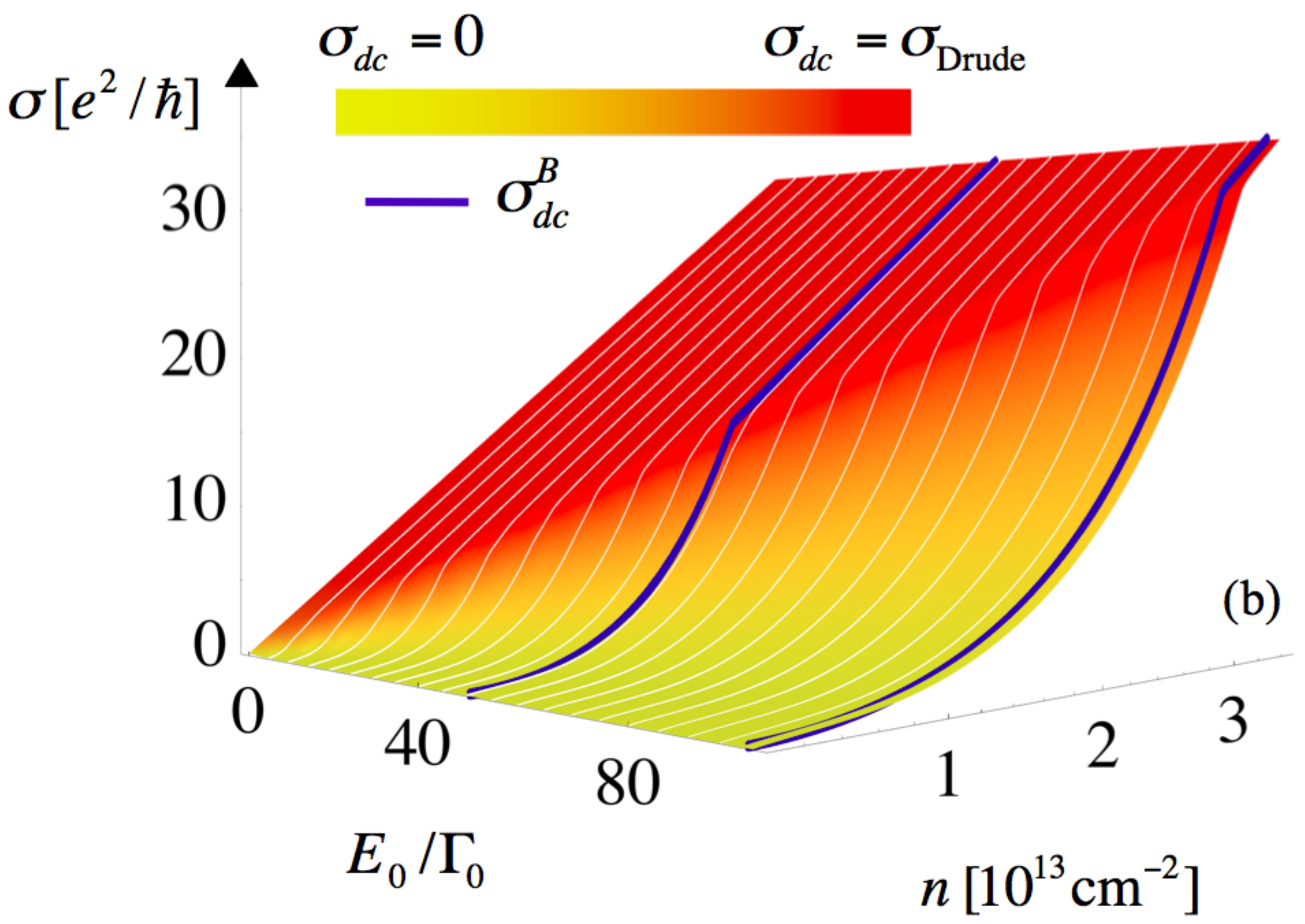}
\end{center}
  \caption{(Color online) (a):
Scattering rate in units of $\Gamma_0$ as a function of  $n/n_0$  for different $E_0$. The dashed line shows $\Gamma_{\rm B}=1/2\tau$
(Eq. \eqref{tauqp}).
(b)Numerical dc conductivity (Eq.~\eqref{sigmaxx}) as a function of $E_0/\Gamma_0$ and $n$ for $\Gamma_0=0.5$meV and  $m=0.7m_e$ (as appropriate e.g. for LaAlO$_3$/SrTiO$_3$ interfaces \cite{biscaras2012}). The  blue lines show Boltzmann conductivity for $E_0=50\,\, {\rm and}\,\, 100\, \Gamma_0$
(c) Mobility $\mu_t$ (solid line) normalized to $\mu^0_t=e/(2\hbar\Gamma_0m)$, compared with Boltzmann's
$\mu_{\rm B}$ (dashed line). Shaded  areas in (a) and (c) denote the low-density regions beyond the diffusive approximation.} 
 \label{fig-sigmamu}
\end{figure}

In conclusion, we have shown that in Rashba 2DEGs  SO coupling entails  an unconventional dc conductivity, strongly dependent on the density and on the SO-coupling strength.
The experimental verification of our results requires the condition
for diffusive transport ($E_F\gg \Gamma$) to be fulfilled in the DSO regime ($E_0>E_F$): this ultimately implies $E_0>E_F\gg\Gamma$.
The conductivity anomalies in the DSO regime can then be accessible
experimentally in relatively clean ($\Gamma_0\sim 1$ meV) samples of
the strong-Rashba  materials mentioned in the Introduction  (see
Refs.\cite{ast2007,ast2008,gierz2009,mirhosseini2010,yaji2010,sanchez2013,gruznev2014,eremeev2012,bahramy2012,sakano2013,xi2013,chen2013,ye2015,xiang2015,gui2004,
  ohtomo2004,reyren2007,caviglia2008,bell2009,benshalom2010,biscaras2012,seri2012,joshua2013,liu2013,hurand2015,zhong2015,caviglia2010,zhong2013,biscaras2014,joshua2012}),
where $E_0\approx 10-140$ meV  and $m\approx 0.2-0.7 m_e$, corresponding to electron densities
$n_0\approx 0.6-8 \times10^{13}$ cm$^{-2}$ \cite{examples}.
Finally, we also remark that the large value of $E_0$ in these systems guarantees that our zero-temperature results will provide a good description for real materials up to temperature scales
$k_B T\sim E_0$. In addition, the occurrence of the DSO anomalies at relatively large densities $n\simeq n_0$ justifies also neglecting of electron-electron interactions, even though a full understanding of the conductivity anomalies in the diluted regime where interactions become relevant is certainly an interesting topic for future investigation.
{\em Acknowledgements}
We gratefully acknowledge fruitful discussions with  S. Caprara,
C. Castellani, M. Grilli and  R. Raimondi. We acknowledge financial
support by Italian MIUR under projects FIRB-HybridNanoDev-RBFR1236VV,
PRIN-RIDEIRON-2012X3YFZ2, Premiali-2012 AB-NANOTECH,
by the European
project FP7-PEOPLE-2013-CIG "LSIE\_2D".

\onecolumngrid
\newpage

\begin{center}
\large{\textbf{Supplemental Material to \\``Unconventional dc Transport in Rashba Electron Gases''}}
\end{center}
\vspace{1cm}
\twocolumngrid
%

\maketitle

\section{Helicity eigenstates basis}
To fix the notation, let us start by giving some details on the helicity eigenstates basis.
The helicity operator is defined as $S=\lf[\hat p\times\vec \sigma\rg]_z$ and its eigenstates, $| \bk s\ra$, with $s=\pm 1$,  satisfy the  relation $S | \bk\, \pm\ra =\pm | \bk\, \pm\ra $.
A simple calculation shows  in particular that $|\bk\, \pm\ra$ can be expressed in terms of the standard spin eigenstates, $|\bk \uparrow\ra$ and $|\bk \downarrow\ra$, as 
$|\bk\, \pm\ra=(\exp{(-i \theta_{\bk})}|\bk \uparrow\ra\pm i |\bk \downarrow\ra)/\sqrt{2}$ with $\theta_\bk=\arctan(k_y/k_x)$.  Consequently,  the matrix $U_\bk$  which implements the rotation from the spin to the helicity eigenstates basis has the form
\be
U_{\bk}=\frac{1}{\sqrt{2}}
\begin{pmatrix}
e^{-i \theta_\bk} & e^{-i \theta_\bk} \\
i & -i \\
\end{pmatrix}.
\ee

In the basis spanned by  the states $| \bk s\ra$ the total Hamiltonian (Eq. (3) in main text) can  be recast as follows:
\be
H=\sum_{\bk} c^\dag_{\bk}\,H_k\, c_{\bk}+
\sum_{\bq,\bk} V_{\rm imp}(\bq) c^\dag_{\bk+\bq} U_{\bk+\bq}^\dag U_\bk c_\bk
\ee
where  $c_\bk=(c_{\bk+},c_{\bk-})$ and $c^\dag_\bk=(c^\dag_{\bk+},c^\dag_{\bk-})$  are spinor creation and annihilation operators,  $H_k$ is the  Hamiltonian of the clean Rashba model, {\sl i.e. } $H_k={\rm diag}(E^+_k-E_0,E^-_k-E_0)$ with  $E^{\pm}_k=(k\pm p_0)^2/(2m)$, $p_0=m\alpha$ and $V_{\rm imp}(\bq)$ denotes the Fourier transform of the impurity potential, $V_{\rm imp}(\bq)=1/{\cal V}\,\sum_j e^{i \bf{q \cdot R}_j} v_{\rm imp}$. 
Differently from the main text, in this Supplementary Material where not differently specified we use units $\hbar=e=1$.
\section{Green's  function}

\begin{figure}[t]
\begin{center}
\includegraphics[width = 1.5 cm]{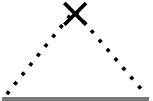}
\caption{Wigwam diagrams which describe the self-energy within Born approximation. The solid line  correspond to the dressed Green function, $G$,  while the crosses indicate averaging over disorder \cite{MahanManyParticlePhysics,Bruus}.}\label{wigwam}
\end{center}
\end{figure}

\begin{figure}[t]
\begin{center}
\includegraphics[width = 7.5 cm]{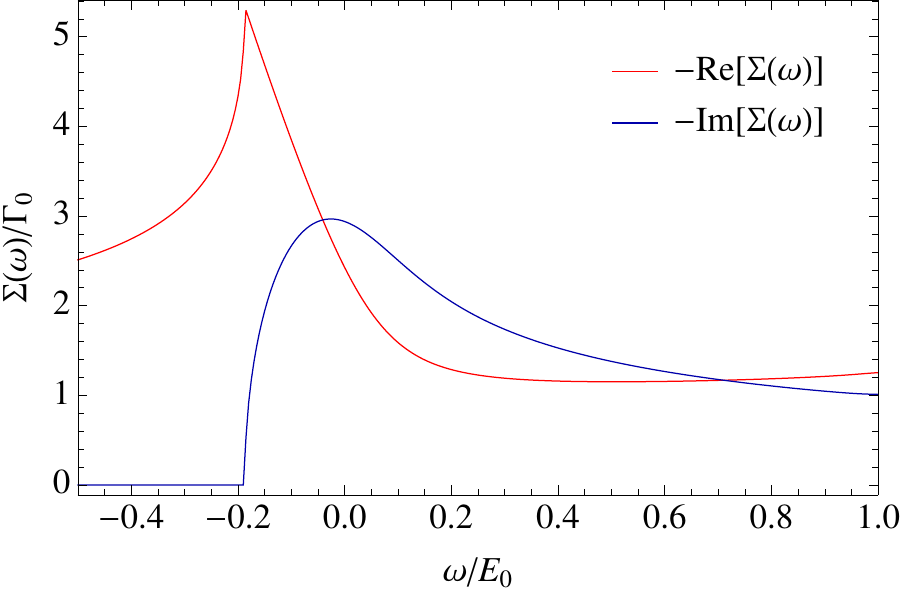}
\caption{Structure of the real and imaginary part of the self-energy close to the lower band-edge for $E_0=40\Gamma_0$.
}\label{resigma}
\end{center}
\end{figure}

The Green's function obeys the standard Dyson equation:
$G^{-1}=\lf(G^0\rg)^{-1}-\Sigma$,  where  $G^0$ is the Green's function of the Rashba model in the absence of disorder and $\Sigma$ is the self-energy.
Specifically, in the helicity eigenstates basis, $G^0$ has the form  
$$[G^0]_{\a\beta}=(i\ve_l-E^\alpha_p+E_F)^{-1}\delta_{\a\b}$$ 
where $\ve_l$ is a fermionic Matsubara frequency, and from now on we set the zero of the energy to $-E_0$. Thus, as in the main text, the dominant spin-orbit (DSO) regime is identified by $E_F<E_0$.  
Within the self-consistent Born approximation (SCBA), $\Sigma=\Sigma({\bf p},i \ve_n)$ is determined by solving the following equation
\begin{equation}
\label{sigmascba}
\Sigma({\bf k},i \epsilon_n)_{}=\frac{n_i v_{\rm imp}^2}{\cal V}
\sum\nolimits_{{\bf p}} U^\dag_{\bf p}U_{\bk}G({\bf p},i\epsilon_n)U^\dag_{\bk}U_{\bf p}
\end{equation}
%
which corresponds 
the ``wigwam diagram''  depicted in Fig. \ref{wigwam} as described {\sl e.g.} in Refs  [\onlinecite{MahanManyParticlePhysics,Bruus}]. Note that in the helicity basis to each impurity-scattering vertex, changing the electron momentum from ${\bf k}$ to ${\bf p}$, one has to associate the spin rotation $U^\dag_{\bf p}U_{\bk}$. 
 Equation \eqref{sigmascba} admits a momentum- and spin- independent solution. Indeed, assuming $
 \lf[\Sigma({\bf p},i \epsilon_n)\rg]_{\alpha\beta}=\Sigma(i \epsilon_n)$, the momentum dependent part on  the r.h.s. of this equation averages away. Within SCBA  the Green's function $G({\bf p},i \epsilon_n )$ is thus represented by the following diagonal matrix in the helicity eigenstates basis:
\be
\label{gren}
\lf[{G }({\bf p},i\ve_l )\rg]_{\a\b}=(i\ve_l-E^\alpha_p+E_F-\Sigma(i \epsilon_l))^{-1}\delta_{\a\b}.
\ee
By analytical continuation to real frequencies of Eq. \eqref{sigmascba} (see {\sl e.g.} [\onlinecite{Bruus}]) we obtain 
the following self-consistent equations for the scattering rate $\Gamma$
\be
\label{gamma}
\Gamma=-{\rm Im}[\Sigma^R(0)]=\frac{n_i v_{\rm imp}^2 \Gamma}{2\cal V} \sum_{\bp} 
\lf[|g_+^R(p,0)|^2+|g_-^R(p,0)|^2\rg]
\ee
and the retarded self-energy $\Sigma^R(\omega)$ 
\be\label{sigma}
\Sigma^R(\o)=\frac{n_i v_{\rm imp}^2}{2\cal V}
\sum\nolimits_{{\bf p}s}g_s^R(p,\o) \theta(p_c-p)
\ee
%
%
where $g_\pm^R(p,\omega)=\left[\o-E^\pm_p+E_F-\Sigma^R(\omega)\right]^{-1}$ is the Green's function of each chiral eigenstate.
To simulate a finite Brillouin zone, in Eq. \eqref{sigma}  we introduced an upper momentum cut-off, $p_c$.
The latter is needed, in particular, to regularize  the real part of the self-energy, 
${\rm Re}[\Sigma^R(\omega)]$, which would otherwise diverge logarithmically at the band edge, see {\sl e.g.} Ref.\cite{knigavko2005}.  
In these regards, we notice that, contrarily to what happens in standard half-filled systems where ${\rm Re}[\Sigma^R(\omega)]$ is approximately $\omega$-independent and it can be absorbed in a redefinition of the Fermi level, in the low-doping regime investigated here ${\rm Re}[\Sigma^R(\omega)]$ acquires a non-trivial frequency dependence. We thus need to  calculate self-consistently both ${\rm Re}[\Sigma^R(\omega)]$ and ${\rm Im}[\Sigma^R(\omega)]$.
Such self-consistent solution identifies the elastic scattering rate,  $\Gamma=-{\rm Im}[\Sigma^R(0)]$ and the renormalized density of states (DOS)
$N(E)=-\frac{1}{\pi{\cal V}}\sum_{\bf p}{\rm Im} [G_R({\bf p},E)]$. The electronic density at $T=0$ is given by  $n=\int_{-\infty}^{E_F} N(E)  dE$.
At low doping the impurities lead to a smearing of the van-Hove singularity in the DOS, which reflects in the behavior of the scattering rate $\Gamma$, as described in the main text (Fig. 3a). In addition, the presence of impurities gives a shift, $\Delta_{\rm edge}$, of the lower  band edge, so that  the lower ``effective'' band edge where $n=0$ is identified by
$\tilde E_F=E_F+\Delta_{\rm edge}=0$.
In Fig.\ref{resigma}, as an example we show the structure of the real and imaginary parts of the self-energy as functions of the frequency, for frequencies close to the lower band-edge, located at $E_F=-\Delta_{\rm edge}$. Notice that these two quantities are connected by  Kramers-K\"onig relations.

%
%

\section{DC conductivity from diagrammatic perturbation theory}
\subsection{Current response function}
Within SCBA  the static conductivity  is given by Eq.(14) of the manuscript, that we report here for convenience
\be\label{sigmaxx}
\sigma_{dc}=\frac{1}{2\pi} \lf(P_{xx}^{AR} -{\rm Re}[P_{xx}^{RR}]\rg).
\ee
The derivation of the above equation in the absence of spin-orbit coupling is standard textbooks material (see {\sl e.g.} Refs. \cite{MahanManyParticlePhysics,Bruus})
and, since it does not change in the presence of spin-orbit coupling, we do not review it here.

\begin{figure}[t]
\begin{center}
\includegraphics[width = 6.5 cm]{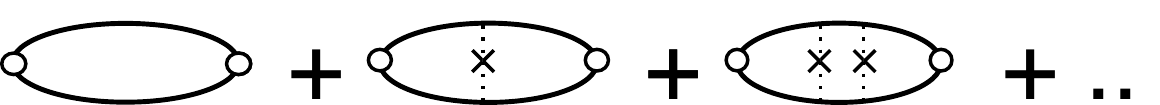}
\caption{Ladder diagrams describing the conductivity within Born approximation. Solid lines and empty circles represent respectively  $G(\bp,i\ve_l)$ and $j_x(\bp)$.}\label{fig-ladder}
\end{center}
\end{figure}

\begin{figure}[b]
\begin{center}
\includegraphics[width = 5 cm]{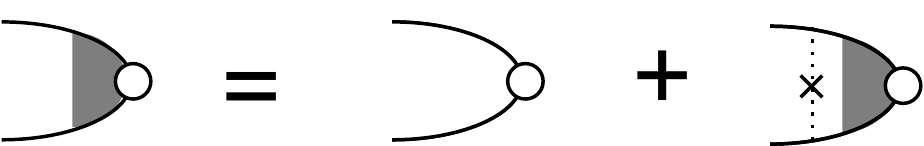}
\caption{Diagrams describing renormalization of the charge current vertex. }\label{fig-svren}
\end{center}
\end{figure}

By applying diagrammatic perturbation theory,  
one easily sees that  the calculation of $\s_{dc}$ implies the summation of all ladder diagrams shown in Fig. \ref{fig-ladder}.
This in turn corresponds to calculate the following current-current response function in  Matsubara  frequencies:
\bea
\label{piba}
P_{xx}(i\ve_{l},i\ve_{l+n})&=&\frac{1}{{\cal V}}\sum_{{\bf p}}{\rm Tr}\left\{
 {G }({\bf p},i\ve_l )j_x({\bf p}){G }({\bf p},i\ve_{l+n}) \right. \cdot \nn \\ 
& \cdot & \left. J_x({\bf p},i\ve_{l},i\ve_{l+n}) \right\}
\eea
where $j_x$ is the bare velocity operator introduced in the main text. We recall that in the helicity basis $j_x=e {\rm v}_x$ is represented by the following matrix:

\be
\label{jbare}
{\rm v}_x=\frac{p_x}{m}\sigma_0+\alpha \cos\theta_{\bf p} \sigma_z +\alpha\sin\theta_{\bf p}\sigma_y
\ee%
where $\sigma_0$ is the $2\times 2$ identity matrix and $\sigma_i$ with $i=x,y,z$ are the Pauli matrices.
The renormalized charge current $J_{x}({\bf p},i\ve_{l},i\ve_m)$ satisfies the diagrammatic equation shown in Fig. \ref{fig-svren} that in helicity space can be written as 

\bea \label{jxkmatsu} 
J_x({\bf k},i\ve_l,i\ve_m)&=&j_x({\bf k})+\frac{n_i v_{\rm imp}^2}{\cal V}\sum_{\bf p}\lf[U^\dag_{\bf k}U_{\bp}{G}({\bf p},i\ve_l)\cdot \rg.\nn\\
& & \lf.J_x({\bf p},i\ve_l,i\ve_m) {G}({\bf p},i\ve_m)U^\dag_{\bf p}U_{\bk}\rg]
\eea
%
%
%
Before coming to the solution of the above equation,  we remark, that, as usual \cite{Bruus}, the AR and RR response functions appearing in Eq.\eqref{sigmaxx} correspond respectively to $P_{xx}(0-i\d,0+i\d)$ and $P_{xx}(0+i\d,0+i\d)$.

As it can be easily verified, by symmetry arguments one finds that $J_x$ has the same matrix structure of the bare current \eqref{jbare}, so that 
we can write:

\be
\label{jdressedmatsu}
J_x(\bk, i\epsilon_l,i\epsilon_m)=
\frac{p_x}{m}\sigma_0+\tilde\alpha \cos\theta \sigma_z +\tilde\alpha\sin\theta\sigma_y
\ee
where $\tilde \a(i\ve_l,i\ve_m)$ satisfies the following self-consistent equation:%
\bea
\label{tildealpha}
\tilde\alpha&=& \alpha+\frac{n_iv_{\rm imp}^2}{2}\sum_{\bf p}{\rm Tr}\big[U^\dag_{\bf p}\sigma_yU_{\bf p}{G}(\mathbf{p},i\ve_l)\cdot \nn\\ & & \cdot J_x(\mathbf{p},i\ve_l,i\ve_m)\,{G}(\mathbf{p},i\ve_m)\big].\eea

Equation \eqref{tildealpha} can be solved explicitly to obtain to the following result for the renormalized anomalous vertex:
\be\label{gcx}
\tilde\alpha(i\ve_l,i\ve_m)=\frac{\a+\a_0(i\ve_l,i\ve_m)}{1-A(i\ve_l,i\ve_m)}
\ee
where we introduced the quantities   $A(i\ve_l,i\ve_m)$ and $\a_0(i\ve_l,i\ve_m)$ given by:
\bea
\!\!\!\!\!\!\!\!\!A(i\ve_l,i\ve_m)&=&\frac{n_iv_{\rm imp}^2}{4 \cal V}\sum_{\bk s s'} {g}_s(k,i\ve_l){g}_{s'}(k,i\ve_m),\label{Aivel}\\
\!\!\!\!\!\!\!\!\!\a_0(i\ve_l,i\ve_m)&=&\frac{n_iv_{\rm imp}^2}{4\cal V}\sum_{{\bf k}s} \frac{k}{m} s\, {g}_s(k,i\ve_l){g}_s(k,i\ve_m).\label{alpha0}
\eea

By replacing Eqs.\eqref{jbare} and (\ref{jdressedmatsu}) in Eq. \eqref{piba}, we  arrive at the following expression for the correlation function $P_{xx}(i\ve_{l},i\ve_m)$,
\be\label{pxxivel}
P_{xx}(i\ve_l,i\ve_m)=P_0(i\ve_l,i\ve_m)+\frac{m\lf[(\alpha+\tilde\alpha)\,\a_0+\alpha\, \tilde\alpha\, A\rg]}{\Gamma_0}.
\ee
On the r.h.s. of the above equation the frequency dependences of   $\tilde\alpha$, $\a_0$ and $A$, 
defined in Eqs.(\ref{tildealpha}-\ref{alpha0}), is implied and we introduced the function $P_0(i\ve_l,i\ve_m)$,
\bea
\label{p0}
	P_0(i\ve_l,i\ve_m)&=&\frac{1}{2\cal V}\sum_{{\bf k}s}\frac{k^2}{m^2}{g}_s(k,i\ve_l){g}_s(k,i\ve_m),
\eea
that  in the absence of Rashba coupling yields the only non-vanishing contribution to the conductivity. 
 
Once the analytical continuation is performed, Eq.\ \eqref{pxxivel} is equivalent to Eqs. (17-18) of the main text. This can be easily seen by inserting in Eq. \eqref{pxxivel} the explicit expression of $A$, $\a_0$ and $P_0$, given  above. One then finds that $P^{LM}_{xx}(0,0)$ can be recast as 
\bea
P^{LM}_{xx}=\frac{1}{2\cal V}\sum_{\bf p}\lf\{\lf[\lf(\frac{p^2}{m^2}+\a\tilde\a^{LM}\rg)(g_+^Lg_+^M+g_-^Lg_-^M)\rg]+\rg.\nn\\ \lf.\frac{p(\a+\tilde\a)}{m}(g_+^Lg_+^M-g_-^Lg_-^M)+\a\tilde\a^{LM}(g_+^Lg_-^M+g_-^Lg_+^M)\rg\}\nn.
\eea
From the above equation we see $P_{xx}$ is the sum of an inter- and intra-band contribution, {\sl i.e.}
\bea
\!P^{LM}_{xx}\!\!&=&\!\!\frac{1}{2\cal V}\sum_{\bf p}\Big\{\sum_s\lf[\lf(\frac{p}{m}+s\a\rg)\!\!
\lf(\frac{p}{m}+s\tilde\a^{LM}\rg)g_s^Lg_s^M\rg]+\nn\\ & & +\a\tilde\a^{LM}\sum_{s\neq s'}g_s^Lg_{s'}^M\Big\}\equiv  P^{LM}_{\rm intra}+P^{LM}_{\rm inter}.\label{pixxlm1}
\eea
Eventually, since $(\frac{p}{m}+s\a)(\frac{p}{m}+s\tilde\a^{LM})\equiv \vec v_{{\bf p}s}\cdot\vec V^{LM}_{{\bf p}s}$ we arrive at
Eq.\ (18) of the main text. 
 
 \subsection{Analytic approximations in the weak-disorder limit}
Starting from the above results in this section we derive approximate analytical expressions for the renormalized vertex and the conductivity.
We assume that we are in the weak-disorder limit (WDL), where we can (i) approximate the spectral functions with a delta, {\sl i.e.} set ${\cal A}_\pm(p,0)=(\Gamma/\pi)|g_+^R(p,0)|^2=\delta(E_\bp^\pm-E_F)$; (ii) neglect the RR contributions.

The WDL approximation for  $\tilde\alpha^{RA}$ can be derived  starting from the analytic continuation of  Eqs.(\ref{gcx}-\ref{alpha0}). 
Performing the angular integral, $A^{RA}$ and $\a_0^{RA}$ can be then cast as 
\be\label{alpha0ra}
\alpha_0^{RA}(0,0)=\frac{n_iv_{\rm imp}^2}{8m\pi}\int_0^\infty \!\!\! p^2 \lf[|g_+^R(p,0)|^2-|g_-^R(p,0)|^2\rg]dp,
\ee
\be
A^{RA}=\frac{n_iv_{\rm imp}^2}{8\pi}\int_0^\infty \!\!\! p\, |g_+^R(p,0)+g_-^R(p,0)|^2dp.
\ee
The latter equation can be simplified using the self-consistent self-energy equation to obtain:
\be\label{ara}
A^{RA}(0,0)=1/2+\frac{n_iv_{\rm imp}^2}{4\pi}\int_0^\infty \!\!\!\! p\, {\rm Re}\!\lf[g_+^R(p,0)\, g_-^R(p,0)\rg]dp.
\ee
Approximating the spectral functions with a delta, as stated above, we immediately see that, except in a small density range around $E_F\simeq E_0$,   we can neglect the second term and the r.h.s. of Eq.\eqref{ara}, since there is no overlap between the two-chiral bands and we obtain 
$A^{RA}(0,0)=1/2$. In the same approximation, setting $\xi^\pm_p=E_p^{\pm}-E_F$, we can write $\alpha_0^{RA}(0,0)$ as follows
\be\label{alpha0ra1}
\alpha_0^{RA}(0,0)=\frac{n_iv_{\rm imp}^2}{8m \Gamma v_F } \int p^2 \lf[\delta(\xi^+_p)-\delta(\xi^-_p)\rg]dp \, ,\ee
which leads to 
%
\be\label{alpha0ra2}
\alpha_0^{RA}(0,0)\simeq-\frac{\Gamma_0 }{4m^2 v_F\Gamma }
\cdot\left\{
\begin{array}{ll}
 (p_-^2-p_+^2)  &\quad E_F>E_0  \\
  (p_+^2+p_-^2)   &\quad E_F<E_0  
\end{array}
\right.
\ee
with $\Gamma_0=n_iv_{\rm imp}^2m/2$.
We recall that here, as in the main text, $p_\pm$ are the momenta on the inner and outer Fermi surface, so that their subscripts refer to the value of the transport helicity $\eta$, introduced in the main text in the context of Boltzmann transport.
The sign change on the r.h.s. of Eq.\eqref{alpha0ra2} is thus due to the fact that for $E_F>E_0$ the
 two contributions come from the two chiral bands in Eq. \eqref{alpha0ra1}, while for $E_F<E_0$ only the $E_p^-$ band contributes, with a two-folded Fermi surface.
  Using the explicit expression $p_\pm$, {\sl i.e.}
\be
\label{ppm}
p^{E_F>E_0}_{\pm}=mv_F\mp p_0\quad {\rm and} \quad p^{E_F<E_0}_{\pm}=p_0\mp mv_F.
\ee  
with $v_F=\sqrt{2E_F/m}$,
we eventually obtain:
\be\label{alpha0ra3}
\alpha_0^{RA}(0,0)\simeq
\left\{
\begin{array}{ll}
-\alpha  &E_F>E_0  \\
 - (E_0+E_F)/p_0 &E_F<E_0  
\end{array}
\right.
\ee
Here we also used the WDL results for $\Gamma$, that can be derived from Eq.\ \eqref{gamma} in the WDL where $|g_\pm^R(p,0)|^2=(\pi/\Gamma)\delta(E_{\bp s}^\pm-E_F)$ as 
\be
\label{gwda}
\Gamma^{WDL}=
\left\{
\begin{array}{ll}
\Gamma_0  &E_F>E_0  \\
 \Gamma_0\sqrt{E_0/E_F}=\Gamma_0 p_0/(mv_F) &E_F<E_0  
\end{array}
\right.
\ee
that coincides with the Boltzmann result from Eq.\ (6) of the main text. 
By replacing the result \eqref{alpha0ra3} into Eq.\ \eqref{gcx}, along with $A^{RA}(0,0)=1/2$, we then obtain the estimate of $\tilde\alpha^{RA}$ quoted in the main text, i.e.
\be
\alpha^{RA}(0,0)\simeq
\left\{
\begin{array}{ll}
0   &E_F>E_0  \\
 \alpha(1-E_F/E_0) &E_F<E_0  
\end{array}
\right.
\label{alpha}
\ee

Let us now discuss the analytic approximation of the  conductivity. As  discussed in main text in the WDL we  can put 
\be
\label{dcapp}
\sigma_{dc}\simeq \frac{P_{\rm intra}^{RA}}{2\pi} \quad ({\rm WDL})
\ee
where the intraband term $P_{\rm intra}^{RA}$ coincides with the first term on the r.h.s. of  Eq. \eqref{pixxlm1}.
By using the result \eqref{alpha} for the anomalous vertex we can rewrite it as follows
 \be\label{condatlast}
 P^{RA}_{\rm intra}=\frac{\lf[p_+^2+p_-^2\rg]}{4m\Gamma} \qquad E_F>E_0\\[0.2cm]
\ee
\be\label{condatlast2}
 P^{RA}_{\rm intra}=\frac{\lf[p_+^2-p_-^2-m\tilde \alpha_{RA}(p_+-p_-)\rg]}{4m\Gamma}\qquad E_F<E_0.
\ee
Using the expression of $p_\pm$, $\Gamma$ and $\tilde\a^{RA}$ derived above, along with the expressions for the particle density in the WDL, i.e. 
\be
\label{nwda}
n=
\left\{
\begin{array}{ll}
(m/\pi)(E_F+E_0) &E_F>E_0  \\
 (p_0^2/\pi)\sqrt{E_F/E_0}=n_0\sqrt{E_F/E_0} &E_F<E_0  
\end{array}
\right.
\ee
Eq.s (\ref{condatlast}-\ref{condatlast2}) lead to the final expression for the conductivity quoted in Eq.s\ (1)-(2) of the main text.

To conclude this section we would like to show that the inclusion of vertex corrections is crucial in both regimes.
The ``bare-bubble'' conductivity $\sigma_{bb}$, corresponding to the first diagram in Fig. \ref{fig-ladder}, is given by the term $P_0$ defined in Eq.\ \eqref{p0}. It can be directly computed from Eq.\ \eqref{pixxlm1} by replacing $\tilde \alpha$ with $\alpha$, so that the renormalized velocity $\vec V_{\bp s}$ is replaced by the bare one $\vec v_{\bp s}$.  In the WDL we then easily obtain
\be
P_{bb}^{RA}=\frac{\pi}{2\Gamma {\cal V}}\sum_{\bp,s}\vec v_{\bp s}^2\delta(E_\bp^s-E_F)=\frac{v_F}{4\Gamma} (p_++p_-)
\ee
Thus, using Eq.s\ \eqref{ppm}, \eqref{gwda} and \eqref{nwda} into Eq.\ \eqref{dcapp} one easily obtains that (restoring the charge $e$)
\bea
\sigma_{\rm bb}&=& \frac{e^2(n-n_0/2)}{2m\Gamma}=\sigma_{\rm Drude}-\sigma_{n_0},  \quad  E_F>E_0  \\
\sigma_{\rm bb}&=& \frac{e^2 n}{4m\Gamma}=\frac{\sigma^{WDL}_{\rm Drude}}{2},  \qquad \qquad  \qquad E_F>E_0  
\eea
where, as in the main text, $\sigma_{\rm Drude}=e^2 n/(2\Gamma_0m)$, $\sigma_{n_0}=e^2n_0/(4\Gamma_0 m)$ and $\sigma^{WDL}_{\rm Drude}=e^2 n/(2\Gamma^{WDL}m)$. 
 We therefore see that the bare-bubble result is inadequate at all densities and chemical potentials. This also shows that even recovering the Drude conductivity at $E_F>E_0$ is a non-trivial result, due to the crucial role of vertex corrections. Indeed, even in the regime  $E_F\gtrsim E_0$, where $E_F$ slightly exceeds the Rashba energy, so that two Fermi surfaces are clearly separated, any signature of the Rashba interaction disappears in the dc conductivity, which is given by the usual Drude formula. We also notice that vertex corrections tend to enhance the conductivity with respect to the bare-bubble result.  As we shall discuss below, this is the result one  usually expects within a Boltzmann picture, where backward and forward scattering processes contribute to the transport scattering time with different weights. 

 \begin{figure}[t]
\begin{flushleft}
\includegraphics[width=8cm]{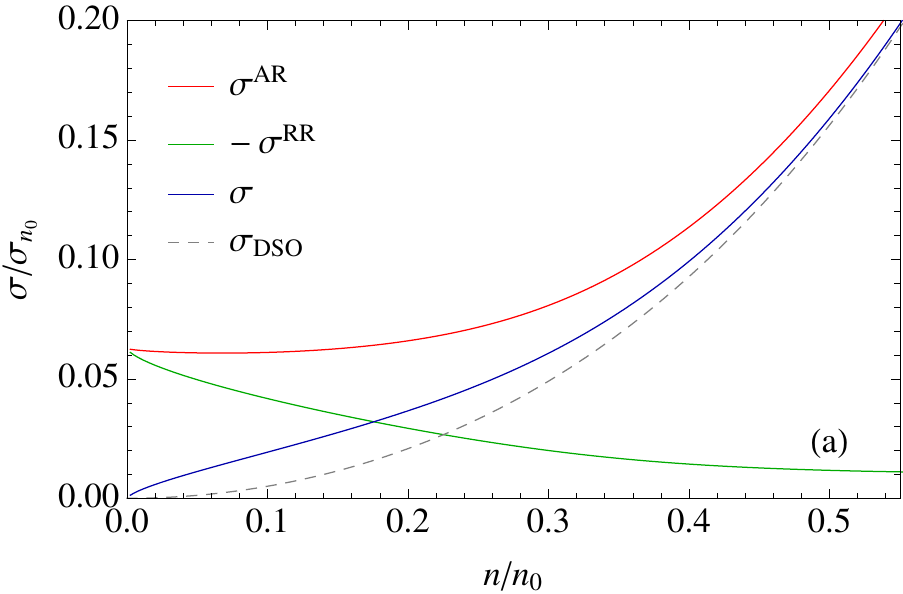}
\includegraphics[width=8cm]{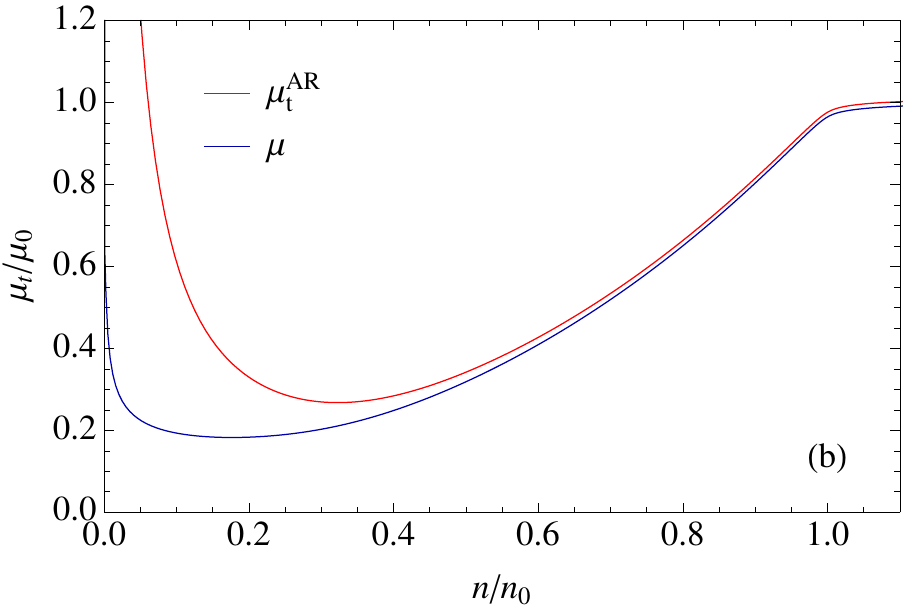}
\end{flushleft}
  \caption{(Color online) (a) Different contributions to the conductivity as a function of the density for $E_0=40 \Gamma_0$ , $\Gamma_0=0.5meV$ and $m=0.7 m_e$. 
  (b) Comparison of the total the $AR$ mobilities. Parameters as in panel (a).} 
 \label{RR-cont}
\end{figure}
 
\subsection{Relevance of the RR contribution}
As we mentioned above, our analytical formulae are in principle valid only in the WDL, realized  for $\Gamma\ll E_F$.
In this limit we did two approximations, we replaced the spectral functions of the chiral eigenstates with  delta functions and neglected the RR part in Eq.\ \eqref{sigmaxx}.
As the density decreases in the DSO regime these approximations are not valid anymore and the quantum result for $\sigma_{dc}$ starts to deviates from the analytical WDL result $\sigma_{DSO}$. In this regime the RR current response function becomes as much relevant as the RA one, and it is important to include it in order to reproduce physical results for the dc conductivity. In Figure \ref{RR-cont}(a) we plot separately the  contributions  of the RR and AR response function to the conductivity,  defined respectively as $\s^{AR}=P^{AR}(0,0)/(2\pi)$ and $\s^{RR}=-P^{RR}(0,0)/(2\pi)$. As one can see, for  our parameters choice, at  $n/n_0\lesssim 0.5$ the RR contribution increases, while the AR one tends to saturates with decreasing density. Neglecting $\sigma_{RR}$ thus leads to the unphysical result of a finite conductivity for zero density. The vanishing of the conductivity as $n\rightarrow 0$ is indeed guaranteed by the cancellation between $\sigma^{AR}$ and $\sigma^{RR}$. 
This fact is also evident in Fig.\ \ref{RR-cont}(b) where along with the total mobility 
 $\mu_t$ we also plot the ``AR'' mobility, $\mu_t^{AR}$ defined as $\mu_t^{AR}=\sigma^{AR}/n$. 
 We se that the inclusion of the RR contribution significantly modifies the structure of the mobility curves around the minimum.
Eventually we note that the cancellation between RR and AR terms  turns out to improve the agreement
between  $\sigma_{dc}$ and  $\sigma_{DSO}$  and it enlarges the range where the DSO formula can be used to describe transport.
\section{Dc conductivity within semiclassical Boltzmann approach}
For a detailed discussion  of Boltzmann equation we refer the reader to Ref.\cite{ziman}, here we only outline the most significant steps.
Note that here we neglect the effect of the off-diagonal terms of the spin-density matrix in the helicity basis at a given ${\bf p}$ 
since as discussed in the main text they give only small corrections to dc charge transport (see e.g. Refs.\cite{dyakonov1984,khaetskii2006,shytov2006,raimondi2006}).
Let us call $\rho_{\bp,s}$ the distribution function for the quasiparticle eigenstates $\ve_{\bp,s}$, where $s$ denotes in general the band index, 
which coincides in our case with the chiral index. The time derivative of $\rho$ is determined by the collision integral:
\be
\lb{coll}
\frac{\pd \rho_{\bp,s}}{\pd t}=-\sum_{\bp',s'} Q^{\bp s}_{\bp' s'}[\rho_{\bp,s}-\rho_{\bp',s'}],
\ee
where $Q^{\bp s}_{\bp' s'}$ is the scattering kernel from the state $\ve_{\bp,s}$ to the state $\ve_{\bp',s'}$.
In the presence of an electric field $\bE$ the l.h.s. of the above equations is given by:
\be\lb{efield}
\frac{\pd \rho_{\bp,s}}{\pd t}=-e \bE\cdot \frac{\pd \rho_{\bp,s}}{\pd \bp}\simeq 
-e \bE\cdot \vec v_{\bp s}
\frac{\pd \rho^{\rm eq}_{\bp,s}}{\pd \ve_{\bp s}}
\ee
in the last passage we replaced $\rho$ with its equilibrium value $ \rho^{\rm eq}_{\bp,s}\equiv f(\ve_{\bp s})$, since we are interested in the  linear response in $\bE$. In the relaxation-time approximation we can express the time evolution of $\rho$ via a transport scattering time $ \tau^{tr}_{p s}$, so that:
\be
\lb{taut}
\frac{\pd \rho_{\bp,s}}{\pd t}=-\frac{\rho_{\bp,s}-\rho^{\rm eq}_{\bp,s}}{ \tau^{tr}_{p s}}
\ee
By combining Eqs.\ \pref{efield}-\pref{taut} we then have:
\be
\lb{gks}
\rho_{\bp,s}=\rho^{\rm eq}_{\bp,s}+e \bE\cdot \vec v_{\bp s}  \tau^{tr}_{p s}\frac{\pd \rho^{\rm eq}_{\bp,s}}{\pd \ve_{\bp s}}.
\ee
For a field in the $x$ direction the current can  then be written as:
\be
j_x=-e\sum_{\bp s} v^x_{\bp s} \rho_{\bp,s}=e^2 E_x \sum_{\bp s} (v^x_{\bp s})^2  \tau^{tr}_{p s} \left(-\frac{\pd \rho^{\rm eq}_{\bp,s}}{\pd \ve_{\bp s}}\right)
\ee
where we used the fact that there is no current in the equilibrium state. 
At $T=0$ we can put $\pd \rho^{\rm eq}_{\bp,s}/\pd \ve_{\bp s}=\pd f(\ve_{\bp s})/\pd \ve_{\bp s}=\delta(\ve_{\bp s}-\mu)$. Thus, by identifying $\ve_{\bp s}-\mu\equiv E_{\bp}^s-E_F$ we arrive at Eq.(8) of the main text%
\be
\lb{sdcb}
\s_{dc}=\frac{e^2}{2}\sum_{\bp s} |\vec v_{\bp s}|^2  \tau^{tr}_{p s} \delta (E_{\bp s}-E_F)
\ee

A set of equations for the transport scattering times can be derived by substitution of Eq.\ \pref{efield} into Eq.\ \pref{coll}, once that one uses the {\sl Ansatz} \pref{taut}.  By doing so, since the scattering kernel $Q$ conserves the energy, and the equilibrium function $\rho^{\rm eq}_{\bp,s}$ does not depend on the chiral index but only on the energy, one is left with:
\bea
\lb{velocity}
\vec v_{\bp s}&=&\sum_{\bp' s'}  Q^{\bp s}_{\bp' s'}\left[  \tau^{tr}_{p s}\vec v_{\bp s}- \tau^{tr}_{p' s'}\vec v_{\bp' s'}\right]=\nn\\ & =& \frac{ \tau^{tr}_{p s}}{ \tau\lf(E_p^s\rg)}\vec v_{\bp s}-\sum_{\bp' s'}  Q^{\bp s}_{\bp' s'} \tau^{tr}_{p' s'}\vec v_{\bp' s'},
\eea
where we introduced the quasiparticle scattering time
\be
\lb{quasi}
\frac{1}{ \tau\lf(E_p^s\rg)}=\sum_{\bp' s'}  Q^{\bp s}_{\bp' s'}.
\ee
By using the fact that only the component of $\vec v_{\bp' s'}$ in the direction of $\vec v_{\bp s}$ survives after momentum integration, Eq.\ \pref{velocity} finally reduces to Eq.(9) of the main text%
\be
\lb{tausol}
\frac{ \tau^{tr}_{p s}}{ \tau\lf(E_p^s\rg)}=1+\sum_{\bp' s'}  Q^{\bp s}_{\bp' s'} \tau^{tr}_{p' s'} \frac{\vec v_{\bp' s'}\cdot \hat v_{\bp s}}{|\vec v_{\bp s}|}
\ee
%
%
%
Notice that  Eq.\ \pref{tausol} differs from the one proposed {\sl e.g.} in Ref.\ \cite{sinova}, where the  band-dependence of the
 transport scattering times on the r.h.s. of Eq.\ \eqref{tausol} has been overlooked, leading to decoupled 
 equations for the $ \tau^{tr}_{p s}$.  Here instead the set of coupled equations \eqref{tausol} is analogous
  to the self-consistence equations \eqref{jxkmatsu} introduced above for the renormalized current in the quantum language. 
  This analogy can be exploited further by the identification of the renormalized Boltzmann current as
\be
\lb{jboltz}
\bJ^B_{\bp s}=e\vec v_{\bp s}\frac{ \tau^{tr}_{p s}}{  \tau\lf(E_p^s\rg)}
\ee
As already discussed in Ref.\ \cite{raimondi_epjb02} for the case $E_F>E_0$,  both the quantum and the Boltzmann approaches lead to the same renormalized currents. This result also holds in the DSO regime $E_F<E_0$, as one can see from the explicit solution for the $\tau$'s derived below.
 

\subsection{Collision integral}
By using Fermi Golden Rule, the scattering rate from the state $|\bp,s\ra$ to the state $|\bp',s\ra$ can be written as:
\be
Q^{\bp s}_{\bp' s'}=\frac{2\pi}{V}  |\langle \bp s |V_{\rm imp}|\bp' s'\rangle|^2 \delta(\ve_{\bp s}-\ve_{\bp's'})
\ee
Using the explicit expression of the helicity eigenstates in  plane waves, we can rewrite the above equation as follows:
\be
Q^{\bp s}_{\bp' s'}=\frac{2\pi}{{\cal V}^2}\lf|\!\int\!\!d{\bf r}\,e^{i(\bp-\bp'){\bf r}} V_{\rm imp}({\bf r})W^{\bp\bp'}_{s's}\rg|^2 \!\!\!\delta(E^s_{\bp }-E^{s'}_{\bp'})
\ee
Here the matrix $\hat W^{\bp\bp'}=U^\dagger_{\bp'}U_{\bp}$ comes from the scalar product of the helicity eigenvectors. 
Within our approximations, (self-averaging delta-correlated disorder and Born scattering), we can write $\langle V_{\rm imp}({\bf r})V_{\rm imp}({\bf r}')\rangle \simeq n_iv_{\rm imp}^2\delta({\bf r}-{\bf r}')$  and we can recast the above equation as
\be
\label{qpp}
Q^{\bp s}_{\bp' s'}=\frac{2\pi}{{\cal V}}n_iv_{\rm imp}^2\lf|W^{\bp\bp'}_{s's}\rg|^2 \!\!\!\delta(E^s_{\bp }-E^{s'}_{\bp'}).
\ee
where 
$$|W^{\bp\bp'}_{s's}|^2=\frac{1+{\rm sign}(ss')\cos (\theta_{\bp}-\theta_{\bp'})}{2},
$$
 so that one recovers Eq. (5) of the main text.
\subsection{Solution of  Boltzmann equations}
 Using the explicit expression of $Q_{{\bf p}s}^{{\bf p'}s'}$, Eq.\pref{tausol} reads
 \bea
\lb{tausola}
\frac{\tau^{tr}_{p s}}{ \tau\lf(E_p^s\rg)}&=&1+\frac{\pi n_i v_{\rm imp}^2}{{\cal V}}\sum_{\bp' s'}(1+s s' \hat p \cdot \hat p') \frac{\vec v_{\bp' s'}\cdot \hat v_{\bp s}}{|\vec v_{\bp s}|}\cdot\nonumber\\
& & \cdot \delta(E^s_{\bp }-E^{s'}_{\bp'}) \tau^{tr}_{p' s'}.
\eea
Now  recalling that $\vec v_{\bp s}=v_{\bp s} \hat p$ and that $|\vec v_{\bp s}|=v_F=\sqrt{2m E_F}$ for $E^s_{\bp }=E_F$, for states at the Fermi level we can rewrite the above equation as:
\be
\lb{tausolb}
\frac{\tau^{tr}_{p s}}{ \tau}=1+\frac{\pi n_i v_{\rm imp}^2}{{\cal V}}\sum_{\bp' s'}(\hat p \cdot \hat p')^2\eta_{p s}  \eta_{p' s'}\delta(E_F-E^{s'}_{\bp'}) \tau^{tr}_{p' s'}
\ee
where we set  $\tau\lf(E_F\rg)=\tau$ and $\eta_{p s}=s(\hat v_{\bp s} \cdot \hat p)=\pm 1 $.
Performing the angular integral ${\bf p'}$ and changing variables from $p,s$ to $E,\eta $ we eventually recover equation (10) of the main text:
\be\label{taueta1}
\frac{\tau_{\eta}^{\rm tr}}{\tau}=1+\frac{1}{4 \tau_0 m v_F}\sum_{\eta'} \eta\eta' p_{\eta'} \tau^{\rm tr}_{\eta'},
\ee
where $\tau_0$ denotes as usual the quasiparticle scattering time in the absence of spin-orbit, $\tau_0=1/(m n_i v_{\rm imp}^2)$,  $p_{\eta}=p_{\eta}(E_F)$ indicate the two Fermi momenta introduced in Eqs.\eqref{ppm} and we set $\tau_{\eta}^{\rm tr}=\tau_{\eta}^{\rm tr}(E_F)$.

To solve this equation it is useful to note that $\tau/\tau_0=mv_F/\bar p_F$ where $\bar p_F=1/2\sum_{\eta}p_{\eta}$ that allows us to recast Eq. \ref{taueta1} as:
\be\label{taueta2}
\bar\tau_{\eta}^{\rm tr}=1+\frac{1}{4}\sum_{\eta'} \eta\eta' \bar p_\eta' \bar \tau_{\eta'}^{\rm tr}.
\ee
with $\bar\tau_{\eta}^{\rm tr}=\tau_{\eta}^{\rm tr}/\tau$ and $\bar p_\eta=p_\eta/\bar p_F$.
As one can easily check, the solution of this equation reads:
$\bar\tau_{\eta}^{\rm tr}=\bar p_\eta$ that coincides with Eq. (12) of the main text.

\end{document}